\documentclass[reprint,showpacs,preprintnumbers,amsmath,amssymb,aip,jap]{revtex4-1}

\usepackage{graphicx}
\usepackage{subfig}
\usepackage{txfonts}
\usepackage{amsmath}
\usepackage{soul}

\begin{document}

\title{\textit{Ab initio} electronic structure calculations of solid, solution-processed metallotetrabenzoporphyrins}

\author{Patrick B. Shea}
\email{pshea@umich.edu}
\homepage{http://www.eecs.umich.edu/omelab}
\author{Jerzy Kanicki}
\affiliation{Organic and Molecular Electronics Laboratory, Department of Electrical Engineering and Computer Science, The University of Michigan, Ann Arbor, MI 48109 USA}



\begin{abstract}
An \textit{ab initio} study of the electronic structures of solid metallotetrabenzoporphyrins (MTBPs) utilized in organic transistors and photovoltaics is presented.  Bandstructures, densities of states, and orbitals are calculated for H$_{2}$, Cu, Ni, and Zn core substitutions of the unit cell of solid TBP, as deposited via soluble precursors that are thermally annealed to produce polycrystalline, semiconducting thin-films.   While the unit cells of the studied MTBPs are nearly isomorphous, substitution of the core atoms alters the structure of the bands around the energy bandgap and the composition of the densities of states.  Cu and Ni core substitutions introduce nearly-dispersionless energy bands near the valence and conduction band edges, respectively, that form acceptor  or deep generation/recombination states.  The following article appeared in J. Appl. Phys. 111, 073709 (2012).
\end{abstract}

\maketitle

\section{Introduction}
Porphyrins serve vital functions in a variety of biochemical applications, including photosynthesis and catalysis, and demonstrate versatile electronic and optical properties.\cite{porphyrin_handbook}  Tetrabenzoporphyrin (C$_{36}$H$_{22}$N$_{4}$, TBP) has been demonstrated as a semiconductor that exhibits thin-film polycrystallinity and field-effect behavior via a soluble precursor molecule, wherein a soluble precursor is spun cast to create an amorphous, insulating film, then thermally annealed to form an insoluble, polycrystalline, semiconducting thin-film.\cite{bp_solution,bp_OFET,bp_overview,pshea_BP_OFET,bp_structure}  The synthetic process used to create the precursor molecules can be altered to create variations of the TBP molecule that exhibit widely different thin-film morphology and electrical performance; H$_{2}$, Cu, Ni, and Zn core substitutions (TBP, CuTBP, NiTBP, and ZnTBP, respectively) in particular have demonstrated organic field-effect transistors (OFETs) with mobilities ($\mu_{FE}$) in the range of 0.01-1 cm$^{2}$/V-s.\cite{pshea_BP_OFET,nitbp_ofets,cutbp_ofets,zntbp_ofet_article}  TBP has also recently been demonstrated as a robust donor material in organic photovoltaics (OPVs).\cite{Ku_TBP_2011,Guide_TBP_2011} 

The bulk of theoretical examinations on the electronic structures of porphyrins have focused on single molecules, typically deriving their interpretation from the Gouterman four-orbital model.\cite{gouterman_porphyrin_substitution,gouterman_porphyrin_assignment,gouterman_porphyrin_spectra}  Many of these single molecule calculations have examined the effects of metallation or peripheral substitution on the electronic structure of the porphyrins.\cite{gouterman_porphyrin_assignment,gouterman_porphyrin_spectra,weiss_porphyrin_calculations,gouterman_porphyrin_substitution,maggiora_porphine_calculations,ziva_tetraaza, lee_porphyrin_spectra,rawlings_porphyrins,rosa_mpcs, shelnutt_substituents,metal_porphryin_structure_bonding,porphine_montecarlo, nguyen_znporph_ground,nguyen_znporph_excited,antipas_states,rovira_fe}  In the solid state, experimental and theoretical studies on the electronic structure and conduction pathways in linearly-stacked metallo- porphyrins and phthalocyanines showed intramolecular charge transport is dependent upon molecular spacing, and inter-stack charge transport dependent on molecular composition.\cite{1D_nitbp_sims,ishikawa_pc,bialek_bandstructure_nipc,orti_1996_solidPC}  The prevalence of copper phthalocyanine (CuPc) in OFETs has led to it being studied by \textit{ab initio} calculations both in the single-molecule and solid forms.\cite{Lozzi_2004_CuPC,Evangelista_2007_CUPC,Aristov_2008_CUPC,Yang_2008_CuPC}  Noguchi \textit{et al.} calculated the crystal growth characteristics of solid TBP, reporting that TBP crystallized along the $\vec{b}$ axis parallel to the substrate plane.\cite{Noguchi_TBP_growth}  With the inherent advantages of precursor-route metallotetrabenzoporphyrins (MTBPs) in OFETs and OPVs, the study and comparison of the electronic structures of solid MTBPs is vital to understanding film behavior and device operation.

In this work, we examine the ground state electronic structures of the unit cells of solid, solution-deposited MTBPs utilized in OFETs and OPVs.  Utilizing the CASTEP implementation of density functional theory (DFT), bandstructures, densities of states, and orbitals are calculated for MTBPs (M=H$_{2}$, Cu, Ni, and Zn).  The solid state bandstructures are compared to the single-molecule results in the literature, and analyzed in light of their applications in OFETs and OPVs.

\begin{table}[b]
\caption{Parameters for CASTEP calculations.}
\label{simulations_settings_table}
\centering
\begin{ruledtabular}
\begin{tabular}{c|c}
Setting & Value\\
\hline
Exchange correlation & LDA \\
Pseudopotential & Norm-conserving \\
Plane-wave basis cutoff & 800 eV\\
\textbf{k}-point sampling & 0.02 \AA$^{-1}$ \\
Scissors operator & 0.0 eV \\
DOS smearing & 0.1 eV \\
\end{tabular}
\end{ruledtabular}
\end{table}
\section{Computational Method }
The theory and implementation of DFT in CASTEP has been described by Segall \textit{et al.}\cite{castep}  The calculations described here were performed using the spin-polarized local density approximation (LDA) with norm-conserving pseudopotentials to describe the exchange correlation potential.  Calculations were also performed with the Perdew-Wang '91 (PW91) implementation of the generalized gradient approximation (GGA) with ultrasoft pseudopotentials and found to agree with the LDA results.  The planewave cutoff energy and \textbf{k}-point set mesh (in a Monhorst-Pack grid) values used in the presented results were selected by examining calculation convergence, and selecting parameter values where changes in the total energy (in terms of eV/atom), and in the energy bandstructure, were negligible with increasing cut-off energy or set density.  Zero net charge was assumed.  The parameters used in the calculations are summarized in Table \ref{simulations_settings_table}.  The bandstructure results are presented as calculated at 0 K, ignoring the band smearing and bandgap narrowing that would occur at elevated temperature; a Gaussian convolution operator of 0.1 eV was applied to the densities of states to approximate such thermal effects.  In this work, no scissors operator is applied to the bandstructure or densities of states, although LDA is known to undercalculate the value of the energy bandgap.\cite{blase2011first}  The results presented are thus compared relative to each other so as to discern trends.  

\begin{figure}
\centering
\includegraphics[width=8.5cm]{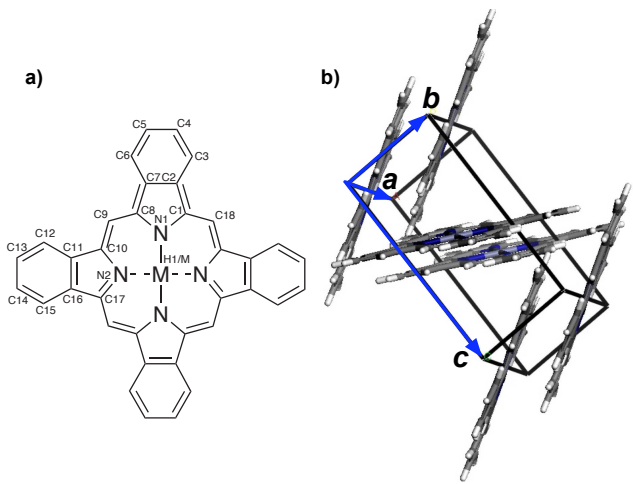}
\caption{(Color online) a) The metallotetrabenzoporphyrin (MTBP) molecule and b)the MTBP monoclinic unit cell.}
\label{molecular_structures}
\end{figure}  
The standard high symmetry notation for primitive monoclinic unit cells was used in this work.\cite{bradley_symmetry}  The crystal structures for the solution-processed MTBPs examined here were determined by powder and thin-film X-ray diffraction (XRD),\cite{bp_structure,nitbp_ofets,cutbp_ofets,zntbp_ofet_article} and were found to be nearly isomorphous.  All unit cells are monoclinic with P2$_{1}$/n symmetry, with the XRD-derived lattice parameters reported in Table \ref{lattice_parameters_table}.  Relevant XRD-derived bond lengths and angles are reported in Table \ref{bond_table} for the solution-processed MTBPs examined here.  Also included in Table \ref{bond_table} are calculated bond angles and lengths for corresponding single-molecule metalloporphyrins.\cite{fleischer_mtpp_structure, fleischer_mp_structure,metal_porphryin_structure_bonding,nguyen_znporph_ground,zhou_2004_photonabsorption,rosa_nickel_spectra}  Prior to energy calculations a geometry optimization step was performed on all unit cells during which the lattice parameters were fixed and the atomic coordinates allowed to relax.  The initial and optimized structures were nearly isomorphous, and their corresponding energy band calculations returned results with no notable difference.  The results presented here are for those prior to geometry optimization.
\begin{table}[b]
\caption{Lattice parameters of reported solid MTBPs derived from X-ray diffraction measurements.  All cells are monoclinic with P2$_{1}$/n symmetry.\cite{bp_structure,cutbp_ofets,nitbp_ofets,zntbp_ofet_article}}
\centering
\begin{ruledtabular}
\begin{tabular}{c|c|c|c|c}
& TBP & CuTBP & NiTBP & ZnTBP\\
\hline
a (nm) & 1.2405  & 1.2390 & 1.2360 & 1.2322 \\
b (nm) & 0.6591& 0.6585 & 0.6578 & 0.6555\\
c (nm) & 1.4927 & 1.5150 & 1.5190 & 1.4994\\
$\beta$ ($^{\circ}$) & 101.445 & 101.160 & 100.620 & 101.748\\
\end{tabular}
\end{ruledtabular}
\label{lattice_parameters_table}
\end{table}

\squeezetable
\begin{table*}
\caption{Relevant bond characteristics for the examined solid MTBPs derived from X-ray diffraction measurements.  Also included are calculated bond characteristics of the single-molecule forms of the respective MTBPs.\cite{zhou_2004_photonabsorption,metal_porphryin_structure_bonding,rosa_nickel_spectra,nguyen_znporph_ground}}
\centering
\begin{ruledtabular}
\begin{tabular}{c||c|c||c|c||c|c||c|c}
& TBP  & SM TBP & CuTBP & SM CuTBP\footnote{Single-molecule form is copper tetraphenlyporphyrin (CuTPP).} & NiTBP & SM NiTBP\footnote{Single-molecule form is nickel porphyrin (NiP).} & ZnTBP & SM ZnTBP \\
\hline
M - N (\AA) & & & 2.040 &  2.029 & 1.990 & 1.972 & 2.089 & 2.078\\
C9 - C10 (\AA) & 1.400 & 1.397 & 1.383 & 1.391 & 1.375 & 1.381& 1.400 & 1.391\\
C11 - C16 (\AA) & 1.424 & 1.423 & 1.415 &  1.365 & 1.409 & 1.362 & 1.425 & 1.411\\
$\alpha$ (M-N-C10) ($^{\circ}$) & 123.5 & 124.7 & 126.4 & 124.5 & 127.2 & 127.8 & 124.2 & 125.8 \\
$\alpha$ (C9-C10-C11) ($^{\circ}$) & 126.0 & 125.7 & 124.0 & 126.9 & 122.9 & 123.7 & 125.9 & 125.1\\
$\alpha$ (N-C10-C9) ($^{\circ}$) & 126.9 & 126.1 & 126.3 & 128.1 & 126.4 & 127.7 & 126.9 & 125.4\\
\end{tabular}
\end{ruledtabular}
\label{bond_table}
\end{table*}

\section{Results and Discussion}
The bandstructures and partial densities of states (PDOS) are displayed in Figures \ref{mtbp_bandstr} and \ref{mtbp_DOS}, respectively, with the corresponding, calculated band properties summarized in Table \ref{bandgaps_table}.  Presentation of the results proceeds as follows.  The bandstructure and partial densities of states (PDOS) are discussed for metal-free TBP.  The analogous results are then presented and compared for the metal-substituted TBPs.  The solid-state results are then compared to the literature on single molecule metallo- porphyrins and phthalocyanines.  In discussion of materials in the solid-state, the nomenclature of valence and conduction bands (VB and CB, respectively) is used.  When discussing single-molecule results or single atoms, the nomenclature of orbitals, such as the highest occupied molecular orbitals (HOMO) and lowest unoccupied molecular orbitals (LUMO) is used.  The top valence bands and bottom conduction bands are given particular attention due to their direct participation in OFET and OPV operation.

\subsection{Bandstructure}
\begin{figure*}
\centering
\subfloat[]{\includegraphics[width=2in]{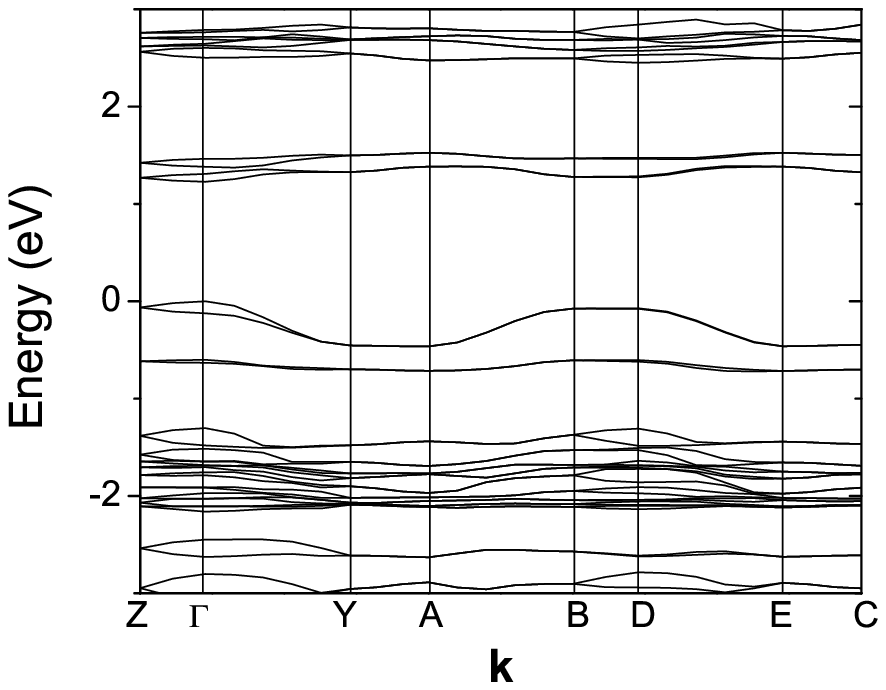}\label{ob6299_bandstructure}}
\subfloat[]{\includegraphics[width=2in]{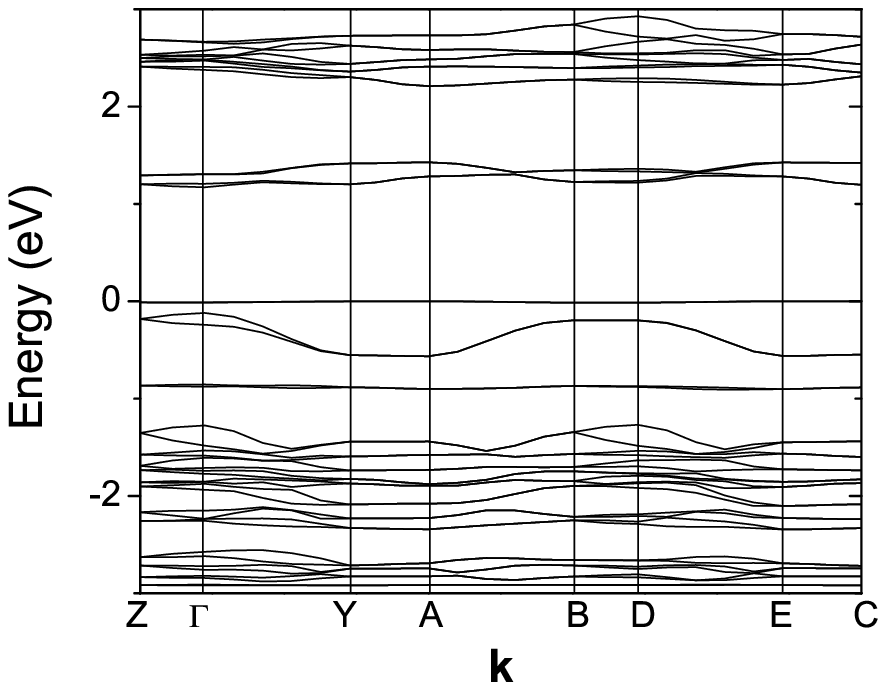}\label{cu_bandstructure}} \\
\subfloat[]{\includegraphics[width=2in]{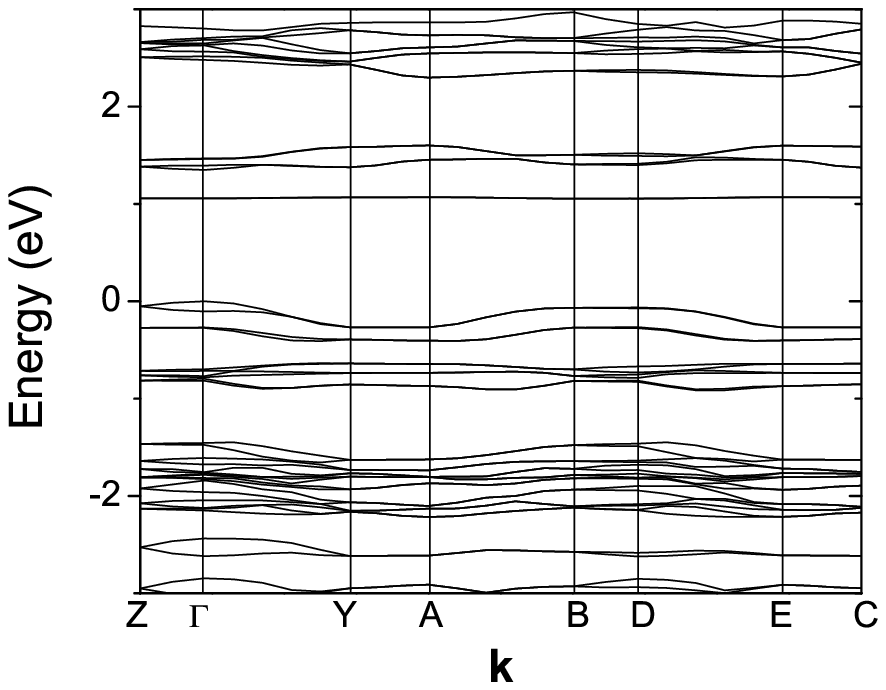}\label{ni_bandstructure}}
\subfloat[]{\includegraphics[width=2in]{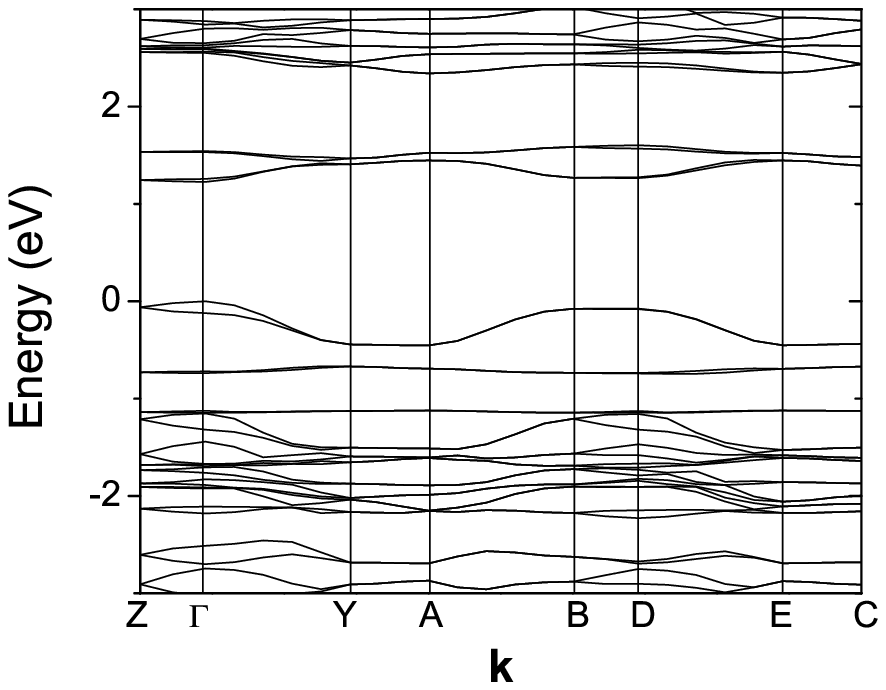}\label{zn_bandstructure}}
\caption[The energy bandstructure of (a) TBP, (b) CuTBP, (c) NiTBP, and (d) ZnTBP.  The 0 eV points represent the top of the respective valence bands.]{The energy bandstructure of \subref{ob6299_bandstructure} TBP, \subref{cu_bandstructure} CuTBP, \subref{ni_bandstructure} NiTBP, and \subref{zn_bandstructure} ZnTBP.  The 0 eV points represent the top of the respective valence bands.}
\label{mtbp_bandstr}
\end{figure*}

\begin{figure*}
\centering
\subfloat[]{\includegraphics[width=2in]{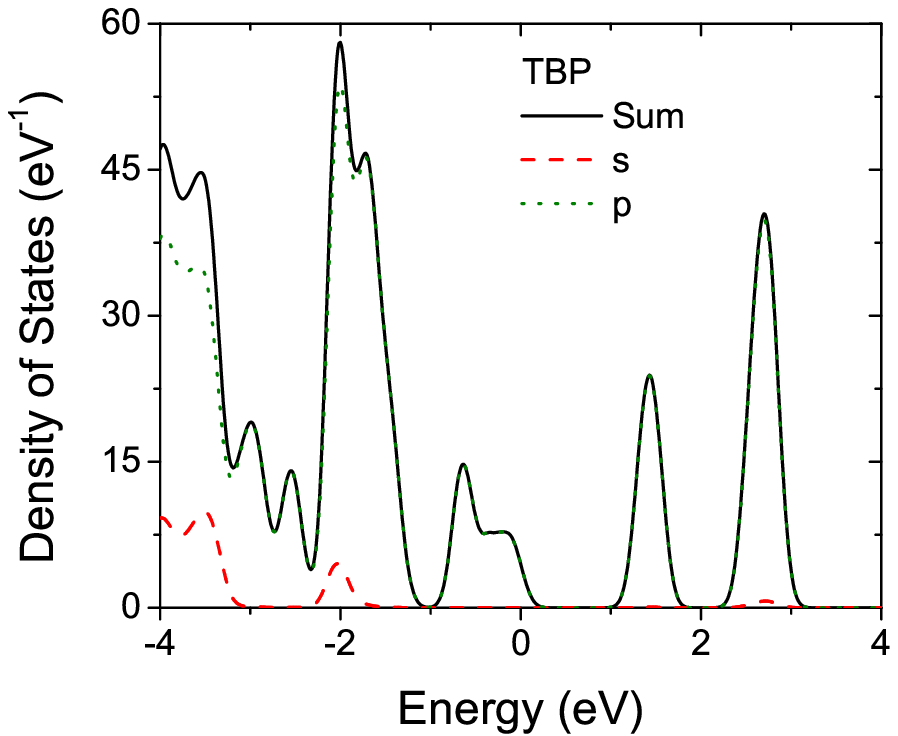}\label{ob6299_DOS}}
\subfloat[]{\includegraphics[width=2in]{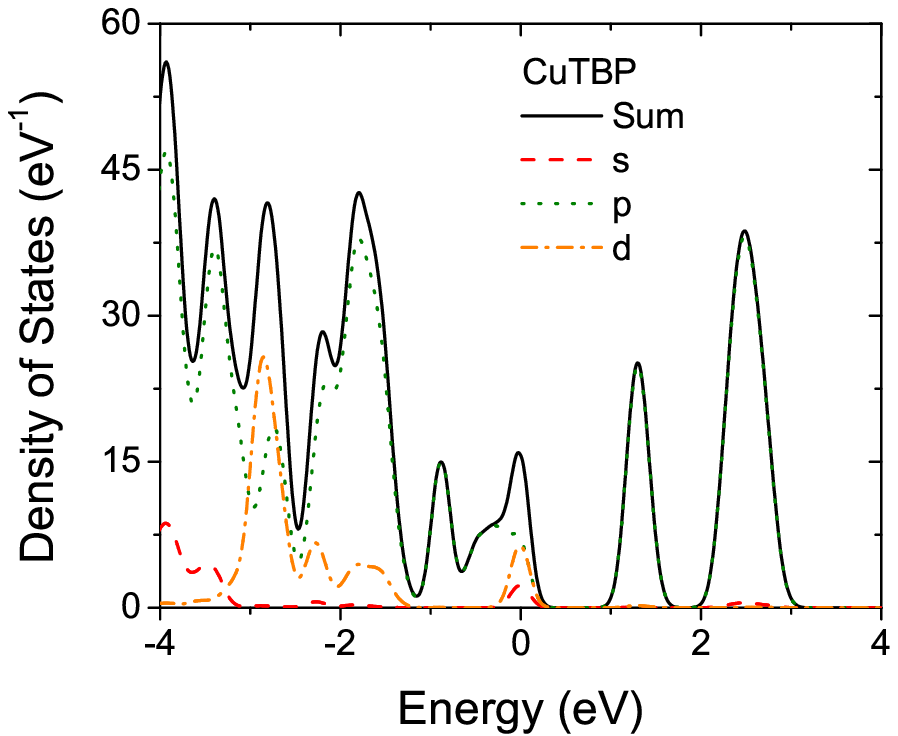}\label{cu_dos}} \\
\subfloat[]{\includegraphics[width=2in]{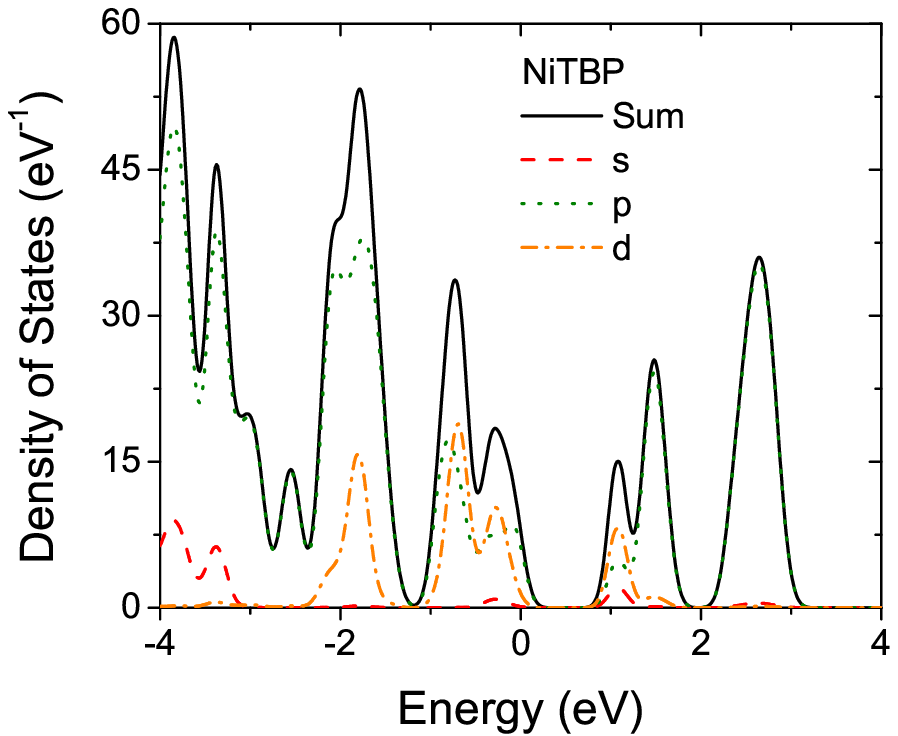}\label{ni_dos}}
\subfloat[]{\includegraphics[width=2in]{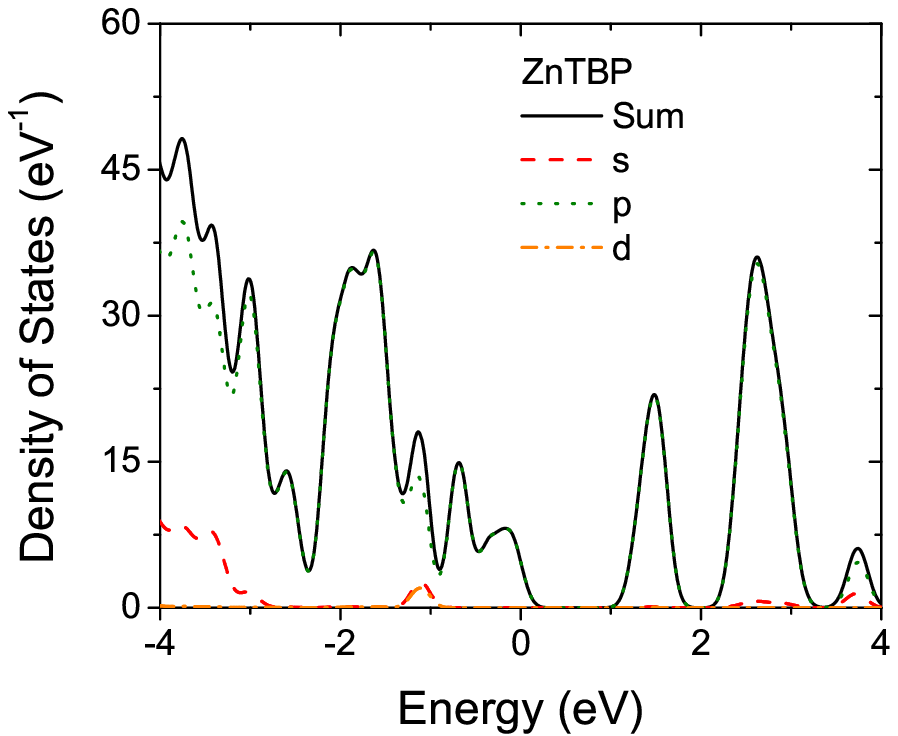}\label{zn_dos}}
\caption[(Color online) The densities of states of (a) TBP, (b) CuTBP, (c) NiTBP, and (d) ZnTBP.  The 0 eV points represent the top of the respective occupied states.]{(Color online) The densities of states of \subref{ob6299_DOS} TBP, \subref{cu_dos} CuTBP, \subref{ni_dos} NiTBP, and \subref{zn_dos} ZnTBP.  The 0 eV points represent the top of the respective valence bands.}
\label{mtbp_DOS}
\end{figure*}

The E vs. \textbf{k} and PDOS diagrams for TBP are shown in Figs. \ref{ob6299_bandstructure} and \ref{ob6299_DOS}, respectively.  Bands appear in pairs due to the presence of two distinct molecules in the unit cell, giving rise to symmetric and antisymmetric molecular wavefunction combinations.  Degeneracy is noted across much of \textbf{k}-space, except between Z-$\Gamma$-Y and B-D-E.  The energy bandgap minimum of E$_{G}$=1.22 eV occurs at \textbf{k}=$\Gamma$.  The top valence band, or valence band maximum (VBM), is composed of 2\textit{p} states from the porphyrin core C atoms, with a smaller contribution from the benzene C atoms.  The bottom conduction band, or conduction band minimum (CBM), is composed of 2\textit{p} states from pyrrole and \textit{meso} C and N atoms, and spreads to the C atoms in the peripheral benzenes, and are parallel to the molecular plane.  The top valence band width (W$_{v}$) are wider than the bottom conduction band widths (W$_{c}$) (W$_{v}$=0.47 eV vs. W$_{c}$=0.16 eV), although for TBP the bottom conduction bands near the bandedge have notably larger densities of states than the bottom valence bands.

The E vs. \textbf{k} and PDOS diagrams for CuTBP are shown in Figs. \ref{cu_bandstructure} and \ref{cu_dos}, respectively.   The VBM and CBM are similar to that of TBP, but of interest is the introduction of a nearly-dispersionless, degenerate band pair above the VBM resulting from the unpaired 3\textit{d}$^{9}$ electron of the Cu$^{2+}$ atom in a 3\textit{d}$_{x^{2}-y^{2}}$ orbital (with significant contribution from the pyrrole N atoms), which in single-molecule and solid CuPc is reported as the singly-occupied molecular orbital (SOMO).\cite{Evangelista_2007_CUPC}  Thus, here the new band is not designated as the VBM due to its composition.  The VBM composition is the same as TBP, and retains the shape of the valence band as observed in TBP.  The CBM demonstrates the same orbital composition as TBP.  The energy bandgap minimum of 1.29 eV occurs at \textbf{k}=$\Gamma$.  However, at \textbf{k}=C the states in the CBM are only 20 meV higher than for \textbf{k}=$\Gamma$.  The difference of less than kT at 300K indicates that as films are utilized at room temperature, thermal effects will narrow the bandgap, smear bands together (see the PDOS in Fig. \ref{cu_dos}), and possibly alter the bandgap from direct to indirect.  The dispersionless SOMO band is a degenerate pair with a maximum occurring at \textbf{k}=E, 0.11 eV above the valence band; the SOMO to conduction band gap is 1.18 eV.  At 300K, this degenerate pair is more than kT away from the VBM.  Like TBP, CuTBP is degenerate except between Z-$\Gamma$-Y and B-D-E.  The band splitting, however, is less than for TBP, particularly in the bottom conduction band.  The top valence bands are wider than the bottom conduction bands, even though the bottom two conduction bands overlap (W$_{v}$=0.44 eV vs. W$_{c}$=0.26 eV).  Furthermore, the bottom conduction band densities of states have a larger sum of states than the top valence bands.

The E vs. \textbf{k} and PDOS diagrams for NiTBP are shown in Figs. \ref{ni_bandstructure} and \ref{ni_dos}, respectively.  Introduction of the Ni$^{2+}$ atom with a 3\textit{d}$^{8}$ orbital induces a nearly-dispersionless, degenerate, unoccupied band pair of 3\textit{d}$_{x^{2}-y^{2}}$ character, but in this case below the CBM (also with significant contribution from the pyrrole N atoms).  As with CuTBP, the VBM and CBM are composed similarly to TBP.  An energy bandgap minimum of 1.35 eV occurs at \textbf{k}=$\Gamma$.  The CBM is shifted upwards relative to the valence band, as compared to TBP.  The  valence band width is more narrow than in the other MTBPs (W$_{v}$=0.27 eV vs. 0.47 eV), while the conduction band width is similar to CuTBP (W$_{c}$=0.25 eV).    However, the minimum of the Ni dispersionless band occurs at \textbf{k}=D, with an energy gap to the VBM of 1.05 eV.  Whereas in CuTBP the 3\textit{d}$^{9}$ orbital contributes notably on the valence band side of the bandgap, in NiTBP the Ni$^{2+}$ 3\textit{d}$^{8}$ orbital contributes on both sides of the band gap.  That is, both the Ni-related mid-gap band, and the band below the VBM comprise contributions by the Ni 3\textit{d}$^{8}$ orbital.  Furthermore, while W$_{v}$ is more narrow in NiTBP, the PDOS reveals that the sum densities of states are notably larger in the VBM than for the other MTBPs.  As observed in CuTBP, degeneracy is more severe for NiTBP than for TBP, with both conduction and valence bands displaying smaller band pair splitting between Z-$\Gamma$-Y and B-D-E.  

The E vs. \textbf{k} and PDOS diagrams for ZnTBP are shown in Figs. \ref{zn_bandstructure} and \ref{zn_dos}, respectively.   Similar structure is noted in the E vs. \textbf{k} diagram compared to TBP, especially in the VBM.  Band splitting is again observed between Z-$\Gamma$-Y and B-D-E.  For ZnTBP, however, the band-pair degeneracy in the CBM is more pronounced than for TBP between Z-$\Gamma$-Y; the bands are also shifted upwards for \textbf{k}$\neq$$\Gamma$.  The energy bandgap minimum of 1.23 occurs at \textbf{k}=$\Gamma$.  The VBM is composed is 2\textit{p} states from the porphin core C atoms, and the CBM is composed of 2\textit{p} states from the porphin core C and N atoms.  Furthermore, the contribution of the Zn 3\textit{d}$^{10}$ orbital near the bandgap is confined to a small number of states 1.12 eV below the top valence band edge.  The valence band width for ZnTBP is still wider than the conduction band width (W$_{v}$=0.45 eV vs. W$_{c}$=0.38 eV), although with a notably smaller difference than in the other MTBPs.  The conduction band width is significantly larger than the other MTBPs, although there is little notable difference in the magnitude of the PDOS.

\subsection{Relation to Single Molecule Calculations}
The wealth of literature on single molecule porphyrins and phthalocyanines illuminates the bandstructure and PDOS results described here for the solid-state.  In Table\ref{bandgaps_table} the HOMO and LUMO orbital energies (E$_{HOMO}$ and E$_{LUMO}$, respectively) of calculated single-molecule metalloporphyrins are presented for comparison to the values calculated for the examined solid MTBPs.  While the results discussed here impact charge transport in thin-film OFETs and OPVs, it is difficult to speculate on OFET and OPV results at present due to the complex nature of those thin films (morphology or contact resistance, for example).\cite{bp_solution,bp_OFET,bp_overview,pshea_BP_OFET,nitbp_ofets,cutbp_ofets,zntbp_ofet_article}  As presented in Table \ref{bond_table}, the measured bond lengths and angles for the solid MTBPs examined here concur with those reported in the single-molecule literature, including the trend in M-N bond length.

The single molecule orbital structure of free-base porphyrins is most commonly associated with the Gouterman four-orbital model,\cite{weiss_porphyrin_calculations,gouterman_porphyrin_substitution,gouterman_porphyrin_assignment,gouterman_porphyrin_spectra} with more recent extensions examining various substitutions or phthalocyanines.\cite{karavaeva_mtbps_spectra,Zhang_2005_dft,Lebedev_2008_tbp,ehrenberg_1990_MTBPs,orti_1996_solidPC,zhou_2004_photonabsorption}  Determination of the HOMO and LUMO orbitals yields similar structure to that observed here in the solid-state.  The D$_{2h}$ symmetry of TBP, versus D$_{4h}$ symmetry in the metal-substituted molecules, is known to lift the molecular orbital degeneracy, as also shown here in the solid-state for \textbf{k} between Z-$\Gamma$-Y.\cite{gouterman_porphyrin_substitution}  Furthermore, Orti \textit{et al.} examined the bandstructure of one-dimensional stacks of similar H$_{2}$Pc molecules and found that notable $\pi$-$\pi$ interactions were exhibited between adjacent molecules, but not between molecular columns, similar to the solid-state structure here demonstrating weak intermolecular interaction.\cite{orti_1996_solidPC}

The popularity of CuPc as an organic semiconductor has produced many experimental and theoretical studies on the electronic structure of its single molecule and solid-state forms.\cite{mpcs_electronc_diffraction_abinitio, metal_porphryin_structure_bonding,bao_mpcs,Lozzi_2004_CuPC,Evangelista_2007_CUPC,Yang_2008_CuPC, Vazquez_2009_CUPC}  The aforementioned SOMO level was noted in solid CuPc, and the HOMO orbitals similarly assigned to the pyrrole C atoms.\cite{Evangelista_2007_CUPC}  The 3\textit{d}$^{9}$ open-shell orbital leaves an unpaired electron in a 3\textit{d}$_{x^{2}-y^{2}}$ orbital with b$_{1g}$ symmetry,\cite{gouterman_porphyrin_spectra,renner_oxidation_cuoetpp} with contributions from the porphin core N 2\textit{p} orbitals.  Yang \textit{et al.} calculated the bandstructure diagram for CuPc; the E$_{G}$ is similar to that reported here for CuTBP, although the degeneracy differs due to the slightly different symmetry.\cite{Yang_2008_CuPC}  The minimal variation from single-molecule to crystalline form indicates that, as shown here, the intermolecular interaction is very weak.\cite{Lozzi_2004_CuPC}

In Ni porphyrins, the Ni$^{2+}$ atom, with its spin-paired, diamagnetic 3\textit{d}$^{8}$ open-shell, is found to also introduce a LUMO orbital of 3\textit{d}$_{x^{2}-y^{2}}$ character.\cite{Krasnikov_2007_Ni,rosa_nickel_spectra, metal_porphryin_structure_bonding}  Krasnikov \textit{et al.} designated the lowest two unoccupied orbitals as LUMO levels, the lowest being the Ni 3\textit{d}$_{x^{2}-y^{2}}$ and the second being related to the porphin C and N 2\textit{p} orbitals.  

Calculations on Zn porphyrins have indicated a closed-shell 3\textit{d}$^{10}$ orbital structure as displayed here.\cite{gouterman_porphyrin_spectra,metal_porphryin_structure_bonding,nguyen_znporph_ground}  Nguyen and Pachter examined several ground state zinc porphyrins, including ZnTBP, and reported similar HOMO and LUMO orbital symmetry (2a$_{1u}$ and 7$_{eg}$, respectively).\cite{nguyen_znporph_ground}  The Zn 3\textit{d} orbital appears in a HOMO-1 orbital (4a$_{2u}$) approximately 0.7 eV below the HOMO level; in our bandstructure the Zn 3\textit{d} orbital appears at 1.1 eV below the valence band.  Furthermore, the solid-state bandstructure displays an intermediate band composed of 2\textit{p} orbitals at 0.7 eV below the valence band.

The calculated single-molecule HOMO and LUMO orbital energies from literature shown in Table \ref{bandgaps_table} illuminate many similar trends to those observed here in solid-state MTBP calculations.  In each case the valence band energy (E$_{v}$) shifts down in energy relative to the single-molecule HOMO energy level.  Similar to the calculated properties for single-molecule metalloporphyrins, metallated MTBPs display a $\pi$-$\pi$* gap wider than the free-base form. Lowering of the E$_{v}$ as displayed in the solid MTBPs was also reported for E$_{HOMO}$ in a series of single-molecule metallated tetraphenylporphyrins.\cite{metal_porphryin_structure_bonding,Liao_2002}  In both the solid-state and single-molecule cases Ni porphyrins display the lowest E$_{v}$/E$_{HOMO}$ and widest band gap.  With respect to the mid-gap states in CuTBP and NiTBP, the E$_{v}$-E$_{SOMO}$ separation in CuTBP is smaller for solid CuTBP than for the single-molecule forms, which suggests a higher degree of Cu-N interaction, while the mid gap band-to-E$_{c}$ separation in solid NiTBP is similar to that reported in single-molecule nickel porphyrins.\cite{metal_porphryin_structure_bonding,Liao_2002,Krasnikov_2007_Ni,rosa_nickel_spectra}

\subsection{Discussion}
The similarities of the energy band structures observed in Fig. \ref{mtbp_bandstr} correspond to the nearly isomorphous crystal structures of the examined unit cells.  In the MTBP bandstructures reported here the degeneracy is indicative of the high degree of symmetry in the molecules (D$_{2h}$ for TBP and D$_{4h}$ for Cu-, Ni-, and ZnTBP) and in the monoclinic unit cell.\cite{gouterman_porphyrin_spectra,ehrenberg_1990_MTBPs}  It is noted in the E vs. \textbf{k} diagrams that the dispersion is very low and the bandgap large between \textbf{k}=Y-A and E-C, indicating planes of least molecular orbital overlap.  In the metal-substituted MTBPs examined here the variation in band-splitting may be indicative of slightly changed molecularly symmetry or planarity, as expected from calculations on single-molecule forms of metalloporphyrins.\cite{nguyen_znporph_ground, metal_porphryin_structure_bonding}  That the band widths are much larger than kT at 300K indicates that only states near the VBM and the CBM are likely to be populated.  However, in less dispersive regions the bands will populate rapidly with increasing temperature, indicating that the band populations will saturate quickly.  The trends in W$_{v}$ and W$_{c}$ imply as well that charge will populate the conduction band states, especially in less dispersive regions, more rapidly than in the valence band states.

\squeezetable
\begin{table*}
\caption{Calculated energy bandgaps (E$_{G}$), valence band energy (E$_{v}$, relative to the vacuum level), valence band widths (W$_{v}$), and conduction band widths (W$_{c}$) of solid MTBPs.  Also included are relevant, calculated single-molecule (SM) HOMO and LUMO energy levels.\cite{zhou_2004_photonabsorption,Liao_2002,nguyen_znporph_ground}}
\begin{ruledtabular}
\begin{tabular}{c|c|c|c|c||c|c}
 Material & E$_{G}$ (eV) & E$_{v}$ (eV) & W$_{v}$ (eV) & W$_{c}$ (eV) & SM E$_{HOMO}$ (eV) & SM E$_{LUMO}$ (eV) \\
\hline
TBP & 1.22 & -5.11 & 0.47 & 0.16 & -4.77 & -2.32 \\
CuTBP\footnote{Single-molecule form is CuTPP.} & 1.29 & -4.96 &0.44&0.26 & -4.89 & -2.37 \\
NiTBP\footnote{Single-molecule form is NiTPP.} & 1.35 & -5.49 &0.27&0.25 & -4.96 & -2.30 \\
ZnTBP & 1.23 & -5.42 &0.46& 0.38 & -4.63 & -2.11 \\
\end{tabular}
\end{ruledtabular}
\label{bandgaps_table}
\end{table*}

In small-molecule films such as pentacene, transistor measurements have shown that the herringbone molecular arrangement, as opposed to a face-to-face arrangement, limits charge transport by increasing intermolecular spacing and reducing $\pi$ orbital overlap.\cite{pentacene_calculations,polycrystalline_grains, ordering_pentacene,functionalized_pentacene_topcontact}  Minari \textit{et al.} examined the effects of intermolecular distance and $\pi$ orbital overlap in a series of metal-substituted octaethylporphyrin OFETs with the molecules arranged face-to-face along the $\vec{c}$ axis, and found $\mu_{FE}$ to be inversely dependent on the intermolecular spacing.\cite{minari2007molecular}  XRD measurements on MTBP thin-films have shown that molecules stack in the herringbone arrangement along the $\vec{b}$ axis, with the molecule face parallel to (114) and the HOMO orbital surface parallel to the molecule face.\cite{Noguchi_TBP_growth}  Thus, in $\vec{a}$ and $\vec{c}$ the intermolecular spacing is small, but $\pi$ orbital overlap is also small.  In $\vec{b}$, however, the intermolecular spacing is small, and the $\pi$ orbital overlap is enhanced.  

That $\vec{b}$ is the preferred path of charge transport is confirmed by examining the curvature of the band diagrams in Fig. \ref{mtbp_bandstr}.  The curvature of the VBM between $\Gamma$-Y indicates the direction of the smallest hole effective mass (m$^{*}_{h}$).  Conversely, the valence band splitting between $\Gamma$-Z (along $\vec{c}$) indicates the lower band displays a positive m$^{*}_{h}$, a reflection of the herringbone alignment of the molecules.  It is also worth noting that the curvature of the VBM in NiTBP corresponds to the lesser W$_{v}$, such that m$^{*}_{h}$ for NiTBP would be expected to be larger than for the other MTBPs.  Furthermore, the curvature of the conduction bands is lesser than for the valence bands near the band gap, indicating that the electron effective mass (m$^{*}_{e}$) is larger than m$^{*}_{h}$.

While the exact nature of the effect of the Cu and Ni 3$\textit{d}$$_{x^{2}-y^{2}}$ orbitals on solid, thin-film charge transport is outside the scope of the analysis performed here, for example due to the widely varying aggregate nature demonstrated in solution-processed MTBP films,\cite{bp_overview,nitbp_ofets,cutbp_ofets,zntbp_ofet_article} given the relative results presented here the nature of those bands could be expected to affect charge transport as follows.  The curvature of the dispersionless, mid-gap bands induced by the Cu$^{2+}$ 3\textit{d}$^{9}$ and Ni$^{2+}$ 3\textit{d}$^{8}$ orbitals in the 3$\textit{d}$$_{x^{2}-y^{2}}$ configuration indicates m$^{*}_{h}$ in the CuTBP band is large, and an electron excited to the NiTBP band will be expected to have a large m$^{*}_{e}$ compared to the CBM.  The partially-filled Cu bands, with rising temperature, accept an electron from the valence band and hence act as a deep acceptor; from the PDOS diagram this acceptor would be expected to contribute charge carriers in an OFET or OPV.  In NiTBP, there are two bands of interest.  The unoccupied mid-gap band, given that it is 0.3 eV from the CBM, potentially behaves as a deep electron trap.  The band below the VBM, on the other hand, would be expected to contribute to charge conduction in an OFET or OPV.  This could potentially explain the higher $\mu_{FE}$ noted in OFETs fabricated from CuTBP and NiTBP ($\geq$0.1 cm$^{2}$/V-s) compared to TBP and ZnTBP (0.01 cm$^{2}$/V-s), although the presented calculations cannot account for grain boundary regions observed in the widely-varying thin-film aggregation.\cite{bp_overview,nitbp_ofets,cutbp_ofets,zntbp_ofet_article}

Charge transport measurements and bandstructure calculations on one-dimensional stacks of metallo- porphyrins and phthalocyanines, including H$_{2}$, Ni, and Zn complexes, have indicated that the interaction of the $\pi$ orbitals from the porphyrin ligand with the metal \textit{d}$_{x^{2}-y^{2}}$ orbital can play a significant role in intramolecular orbital formation and intermolecular charge transport along the one-dimensional stack and in bulk crystals.\cite{1D_nitbp_sims,ishikawa_pc,orti_1996_solidPC,bialek_bandstructure_nipc} Ishikawa, for example, found by calculation and solution absorption spectroscopy that linear stacks of closed-shell phthalocyanines could be treated as metal-free phthalocyanines.\cite{ishikawa_pc}  Comparatively, whereas conduction in one-dimensional stacks of NiPc has been suggested to result solely from ligand orbitals,\cite{schramm_1980_nipc} conduction in one-dimensional NiTBP has been shown to involve both the ligand $\pi$ orbitals and the metal \textit{d} orbitals.\cite{martinsen_1982_nitbp}  The limiting factor in charge transport within the unit cell can thus be expected to be the $\pi$ orbital overlap resulting from the herringbone molecular packing.  Limited enhancement to charge transport can be expected from 3\textit{d} orbital introduction, even with the introduction of charge-contributing acceptors, without improvement to molecular packing, such as in the $\mu_{FE}$ enhancement in OFETs observed in TIPS-pentacene compared to pentacene.\cite{functionalized_pentacene_topcontact,park_2007_tips,minari2007molecular}

\section{Conclusions}
The energy bandstructures and densities of states of solid metallotetrabenzoporphyrins used in solution-processed OFETs and OPVs have been investigated and analyzed.  Calculations indicate the effect of metallation depends on the orbital structure of the substituent, such that metals with open-shell 3\textit{d} orbitals (Cu and Ni) induce interactions between the 3\textit{d} states of the metal and $\pi$ states of the tetrabenzoporphyrin molecule near the bandgap, potentially acting as acceptor states in Cu and deep trap states in Ni.  An empty core (H$_{2}$) or one with a full 3\textit{d} orbital (Zn) are similar, with negligible contribution of \textit{d} states near the bandgap.  Otherwise the structures of the unit cells are nearly isomorphous and isoelectronic.  Furthermore, the bandstructure diagrams indicate enhancement is limited by weak intermolecular interaction.

\begin{acknowledgments}
The authors thank Dr. A. Nwankpa and Messrs. A. Caird and B. Palen in the Center for Advanced Computing at the University of Michigan for invaluable assistance when this work was originally performed in 2006.
\end{acknowledgments}

%

\end{document}